\documentclass[journal]{article}                                    

\usepackage{arxiv}
\usepackage{hyperref}
\usepackage{graphicx}          
\usepackage{amsmath} 
\usepackage{amssymb}  

\usepackage{amsthm}
\usepackage{color,soul}
\usepackage{flushend}
\usepackage{cite}
\usepackage{mathtools}
\usepackage{xcolor}
\usepackage{array,multirow}
\usepackage{algorithm} 
\usepackage{algorithmic}  
\usepackage[linesnumbered,algo2e,ruled,vlined,norelsize]{algorithm2e} 

\allowdisplaybreaks
\newtheorem{FF}{Residual Feature}

\usepackage{authblk}
\title{\LARGE \bf
Koopman Mode-Based Detection of Internal Short Circuits in \\Lithium-ion Battery Pack
} 

\author[1]{Sanchita Ghosh}
\author[1]{Soumyoraj Mallick}
\author[1]{Tanushree Roy}
\affil[1]{Department of  Mechanical Engineering, Texas Tech University, Lubbock, TX 79409, US. Emails:~{\tt\small sancghos@ttu.edu, somallic@ttu.edu,  tanushree.roy@ttu.edu}.}
\begin{document}

\maketitle
\pagestyle{empty}

\begin{abstract}
Monitoring of internal short circuit (ISC) in Lithium-ion battery packs is imperative to safe operations, optimal performance, and extension of pack life. Since  ISC in one of the modules inside a battery pack can eventually lead to thermal runaway, it is crucial to detect its early onset. However, the inaccuracy and aging variability of battery models and the unavailability of adequate ISC datasets pose several challenges for both model-based and data-driven approaches.  Thus, in this paper, we proposed a model-free Koopman Mode-based module-level ISC detection algorithm for battery packs. The algorithm adopts two parallel Koopman mode generation schemes with the Arnoldi algorithm to capture the Kullback-Leibler divergence-based distributional deviations in Koopman mode statistics in the presence of ISC. Our proposed algorithm utilizes module-level voltage measurements to accurately identify the shorted battery module of the pack without using specific battery models or pre-training with historical battery data.  Furthermore, we presented two case studies on shorted battery module detection under both resting and charging conditions. The simulation results illustrated the sensitivity of the proposed algorithm toward ISC and the robustness against measurement noise. 

\end{abstract}

\section{Introduction}

Short circuits are the primary cause behind delayed fire in damaged battery packs or thermal runaway of Li-ion battery systems \cite{zhang2021internal}. In particular, ISCs in Li-ion batteries may occur due to several reasons which include dendrite growth, separator tearing, or electrolyte leakage inside a cell which in turn can lead to internal temperature rise followed by voltage drop within that cell \cite{mallick2023thermal, huang2021review}. Timely detection of ISC is thus essential for the safe operation of batteries.



This goal has motivated  active research in the area of ISC detection and estimation and this research can be categorized into  model-based approaches and data driven approaches. Model based detection of ISCs have been achieved by estimation of  model parameters such as short circuit resistance and current by incorporating the equivalent circuit model (ECM) \cite{chen2016model, sun2021modified}.  Short circuit resistance has typically been estimated with open circuit voltage (OCV) as a function of state of charge (SOC) and internal resistance parallelly obtaining the voltage drop of the model \cite{feng2018detecting, feng2016online}.


(i) \textit{Model-based approaches:} Specifically, techniques such as recursive least square (RLS) algorithm \cite{seo2020online}, ellipsoidal observer in conjunction with Cholesky factorization \cite{xu2022online}, remaining charging capacity based prediction \cite{kong2018fault},  $H_\infty$ nonlinear observer \cite{xu2022soft}, shorting current estimation using Lyapunov based observer \cite{bhaskar2024post} have been proposed. Also, such ISCs can be estimated at multi-scale level by utilizing a number of multi-scale short circuit resistance estimation methods \cite{xu2020multi}. Additionally, the effects of diffusion voltage, additional decrease in the OCV induced by the self-discharge current in a faulty cell, has been studied with the short circuit characteristics \cite{seo2018detection}. In \cite{moon2024short}, authors proposed a joint Moving Horizon Estimation (MHE) approach to simultaneously estimate the short circuit current and the battery capacity and illustrated validation with experimental data. However, these techniques fail to fully address battery aging,  model uncertainties, and parameter variations from cell to cell. As a result, the estimated parameters can be inaccurate, leading to the unreliable detection of ISC using ECM model-based techniques \cite{alqahtani2024parameters, fogelquist2023error}. Hence, it is necessary to look for algorithms that are robust to these uncertainties and variations to estimate the ISCs in battery packs.

To address this, a number of researchers have preferred using Kalman Filter (KF) based approaches. It can be observed that a variety of RLS based methods have been proposed with variant forgetting have been proposed to compensate the effect of noise-corruptive data acquisition in order to estimate short circuit resistance \cite{hu2021disturbance}. Similarly, incorporation of dual extended KF on separated time scales improved estimation accuracy by mitigating noise while estimating short circuit current \cite{yang2023internal}, addition of the RLS with cell difference model through EKF helped to identify the extra depleting current and parallely diagnosing the short circuit \cite{gao2018micro}, incorporation of EKF and coulomb counting method has been used to estimate SOC and to detect short circuit by computing the difference \cite{yang2020board}.  Nevertheless, KF algorithms rely primarily on linear approximation of state or measurement equations and often underestimate system uncertainties. These errors lead to low detection or ISC estimation accuracy and larger convergence time \cite{moon2024short}.  Therefore, the model-based approaches fail to simultaneously address the robust performances against uncertainty or system variability and improved computational efficiency.

\textit{(ii) Data-driven approaches:} Several attempts have been made to determine ISC through data-driven approaches. For instance, extreme learning machine-based thermal (ELMT) model was explored to study battery temperature fluctuations under short circuit conditions \cite{yang2021extreme}, Pearson Correlation Coefficient (PCC) has been introduced to detect ISC for real-time applications \cite{ahmadzadeh2024real}, Principal Component Analysis (PCA) has been incorporated to detect ISC from over-discharged battery pack \cite{schmid2021online}, and random forest classification has been used for short circuit detection \cite{naha2020internal}. Moreover,  dynamic time warping has been implemented for accurate detection and location of ISC cells \cite{qiao2024data}. Since performing experiments on Li-ion batteries to generate ISC data is resource-intensive, finding open-source and reliable battery datasets for battery short circuits can be challenging. This implies that the available sources are often inadequate for training data-driven models, frequently leading to inaccurate results \cite{wang2023model}.

To address these research gaps, our contribution in this paper lies in proposing a Koopman mode-based ISC detection algorithm to identify the shorted battery modules in a Li-ion battery pack. The algorithm adopts an online framework that requires only the module voltage measurements and trains online with limited data without any prior knowledge of the battery model. Thus, the algorithm is inherently generalizable and can be readily implemented for different battery packs with diverse physio-chemical characteristics and various operating conditions.  Additionally, the algorithm is robust against system uncertainties and measurement noise, while being sensitive towards ISC. Finally, we present two simulation case studies for a battery pack with ISC under resting and charging conditions, to illustrate the efficacy of our proposed algorithm.

The rest of the paper is organized as follows: Section~II introduces the preliminaries on Koopman Operator theory and the Arnoldi algorithm. In Section~III, we presented the ISC detection methodology and the detection algorithm. Section~IV presents simulation results, demonstrating the successful detection of ISC in resting and charging batteries. Finally, Section~V concludes the paper and outlines future research directions.







\section{Preliminaries} \label{prob}
In this section, we presented a brief review of the Koopman operator and the utilization of Arnoldi algorithm to approximate Koopman modes.
\subsection{Koopman operator theory}
The Koopman operator (KO) is an infinite-dimensional linear operator that can advance observable functions of the states of a finite-dimensional non-linear dynamical system \cite{brunton2019notes}. To describe the details of the KO, let us first consider the nonlinear dynamics of a Li-ion battery pack defined as:
\begin{align}
    x_{k+1} &= f(x_k); \qquad y_k = h(x_k), \label{fx}
\end{align}
where $x_k \in  \mathbb{R}^d$ is the battery state vector at $k^{th}$ instant and $f: \mathbb{R}^d \rightarrow \mathbb{R}^d$ is a continuously differentiable nonlinear function that captures the state dynamics.    $h : \mathbb{R}^d \rightarrow \mathbb{R}^q$ denotes the nonlinear output function and  $y_k \in \mathbb{R}^q$ is the output at $k^{th}$ instant that contains the  module voltages of the battery pack.  Next, let us consider scalar-valued observable functions  $\psi: \mathbb{R}^d \rightarrow \mathbb{C}$ such that  $\psi\in \mathcal{F}$ and $\mathcal{F}$ is an infinite-dimensional Hilbert space of observable functions. Then, on this space of observables, KO $\kappa : \mathcal{F} \rightarrow \mathcal{F}$ is defined as
\begin{align}
    \kappa \psi (x) = \psi (f(x)). \label{psi}
\end{align}
Additionally, for eigen-observable functions $\phi \in \mathcal{F}$ with corresponding eigenvalues $\lambda \in \mathbb{C}$ of the KO $\kappa$ satisfy: 
\begin{align}
    \kappa \phi = \lambda \phi. \label{eigen}
\end{align}
$\phi(x)$ and $\lambda$ are respectively referred to as Koopman eigenfunction and Koopman eigenvalues. Now, from \eqref{psi}-\eqref{eigen} it yields 
\begin{align}
    \phi(x_{k+1}) = \kappa \phi(x_k) = \lambda \phi(x_k). \label{philambda}
\end{align}
The significance of \eqref{philambda} lies in the fact that the KO $\kappa$ linearly evolves the Koopman eigenfunction $\phi$ in time. Now, for any other observable function $\psi(x)$ that lies on the space $\mathcal{F} = \text{span}\{\phi_i\}_{i=1}^\infty$, can be expanded as
$\psi(x) = \sum\limits_{i=1}^\infty \phi_i(x) v_i^\psi$. Here, $v_i^\psi$ are the coefficient of the projection of $\psi(x)$ onto the $ \text{span}\{\phi_i\}_{i=1}^\infty$ and are referred as Koopman Modes (KMs). Such expansion of observable functions in terms of Koopman eigenfunctions and KMs is referred as Koopman Mode Decomposition (KMD) \cite{ghosh2024koopman}. Moreover, applied Koopman theory focuses on obtaining the finite subset of $\phi(x)$ such that if $\psi(x)$ lies on the span of this finite subset of eigenfunctions, i.\,e., $ \text{span}\{\phi_i\}_{i=1}^n$, then we can obtain $\psi(x) \approx \sum\limits_{i=1}^n \phi_i(x) v_i^\psi$. Now, we can assume that there exists a finite subset of eigenfuctions such that the output function of the system \eqref{fx} that lies onto the $ \text{span}\{\phi_i\}_{i=1}^n$, and thus, we can obtain,
\begin{align}
    y_k = h(x_k)  = \sum\limits_{i =1}^n \lambda_i^k \phi_i(x_0) v_i^h = \kappa^k h(x_0)= \kappa^k y_0 \label{KMD}.
\end{align}
\noindent
Several data-driven approaches 
can be incorporated to find the finite subset of the eigenfunctions and thus estimate finite sum approximation of KO in \eqref{KMD} \cite{brunton2019notes}. In this paper, we focus on Delay Embedding, often quoted as Hankel Dynamic Mode Decomposition (HDMD). HDMD is a data-driven approach where Taken's theorem \cite{sauer1991embedology} is exploited to obtain a reliable approximation of the KO  from a consecutive measurement data sequence \cite{ghosh2024koopman}. 
The delay embedding approach is often referred to as . To implement this method, we arrange the available consecutive measurement data $Y_\mathcal{L} = \begin{bmatrix}
y_k & y_{k+1} & \cdots & y_{l}
\end{bmatrix}$ over a time period $\mathcal{L}$ as follows. 
\begin{align} 
    &\Upsilon_o = \begin{bmatrix}
        (\overline{\Upsilon}_1) & (\overline{\Upsilon}_2) & \cdots & (\overline{\Upsilon}_{l-\tau-1})
    \end{bmatrix},  \label{win1} \\
    &\Upsilon_u = \begin{bmatrix}
        (\overline{\Upsilon}_2) & (\overline{\Upsilon}_3) & \cdots & (\overline{\Upsilon}_{l-\tau})
    \end{bmatrix}. \label{win2}
\end{align} Here $\overline{\Upsilon}_l = \begin{bmatrix}
    y_k^T & y_{k+1}^T  & \cdots & y_{k+\tau}^T
\end{bmatrix}^T$, where $y_k $ is  defined in \eqref{fx} and $\tau$ is the embedded delay. Finally, the approximate KO can be obtained as,
\begin{align}
    \kappa = \Upsilon_u \Upsilon_o^\dag. \label{KOapp}
\end{align}
We primarily use the HDMD to approximate Koopman linear model over a learning window. Next, we use the Arnoldi algorithm to generate Koopman modes over a prediction window. The next section describes the Arnoldi algorithm.
\subsection{Arnoldi-based Koopman mode generation}
A variant of standard Arnoldi's algorithm was proposed to approximate KMs in \cite{rowley2009spectral} by considering a time-series measurement data $Y$
 under uniform sampling. 
 From \eqref{KMD}, we have the linear approximate model as $y_{k+1} = \kappa y_k $. Now, the first $k$ time-series data of $Y$ spans the Krylov subspace $\text{span}\{\Theta\}, \, \text{where} \, \Theta \coloneqq \begin{bmatrix}
     y_0 & y_1 & \cdots & y_{k-1}
 \end{bmatrix}$. Then, the eigenvalues  of the KO can be approximated by projecting $\kappa$ onto the subspace $\text{span}\{\Theta\}$. These approximated eigenvalues are called the Ritz values. Furthermore,  the approximated Ritz vectors corresponding to the Ritz values behave in the same manner as  $\phi_i(x_0)v_i^h$, i.\,e., KMs scaled by   $\phi_i(x_0)$. 
 
Now, we present the steps of the Arnoldi algorithm. Let us define \mbox{$\mathbf{a} = \begin{bmatrix} 
a_0 & a_1 & \cdots & a_{k-1}
\end{bmatrix}^T$} to  approximate $y_k$ as a linear combination of the previous $k$ time-series data, 
          such that $ y_k = \Theta \mathbf{a}  - e$.
 The approximation error $e \in \mathbb{R}^q$ can be minimized when $\mathbf{a}$ is selected such that \mbox{$e \perp \text{span}\{\Theta\}$}. Let us define the companion matrix  $C = \begin{bmatrix}
     \textbf{0} & a_0 \\ I_{k-1} & \overline{a}
 \end{bmatrix}$, where $\overline{a} = \mathbf{a} \setminus a_0$. Now, since $y_{k+1} = \kappa y_k $, we can write  $\kappa \Theta = \Theta C$ for $e = 0$. 
 In this case, for any $ C v_c = \lambda_c v_c$, and $v_a = \Theta v_c$, it can be shown that $\kappa v_a = \lambda_c v_a$.  Hence, the eigenvalues of $C$ are a subset of eigenvaules of $\kappa$, i.\,e., $\lambda_c \subset \lambda_i$.  However, in the case of $e \neq 0$,
 the eigenvalue $\lambda_c$ of the companion matrix $C$ will be the approximated Ritz values of the system corresponding to the approximated Ritz vectors $v_a = \Theta v_c$. Furthermore, for Ritz values $\Lambda_i = [ \lambda_{c_1}^i, \cdots, \lambda_{c_k}^i]$, 
the corresponding Ritz vectors $\overline{v}$  can be obtained as
\begin{align} \label{vand}
   \overline{v} = \Theta T^{-1}, \quad T = \begin{bmatrix}
   \textbf{1}^T & \Lambda_1^T & \Lambda_2^T & \cdots & \Lambda_{k-1}^T
\end{bmatrix}.
\end{align}
Here,  $ C = T^{-1} \text{diag} (\Lambda_1) T$, and the columns of  the Vandermonde matrix $T$  are eigenvectors $v_c$.

At the same time, using the KO $\kappa$, we can obtain \mbox{$\kappa \Theta = \Theta C + \overline{e}$}, where $\overline{e} = \begin{bmatrix}
    0 & \cdots & 0 & e
\end{bmatrix}^T$. Then,
\begin{align}
     \kappa \Theta T^{-1} = \Theta T^{-1} \Lambda + \overline{e} T^{-1} \,\, \Rightarrow \,\, \kappa \overline{v}\, = \,\overline{v} \Lambda + \overline{e} T^{-1}.
\end{align}
Therefore,  the Ritz parameters approximate the Koopman parameters such that Ritz values $\Lambda_1$ are the approximate Koopman eigenvalues $\lambda_i$ and Ritz vectors $\overline{v}$ are the Koopman modes $v_i^h$ scaled by the initial condition $\phi(x_0)$. 
With these preliminaries on KO and Arnoldi-based KM generation, we now present our proposed ISC detection algorithm.

\section{ISC Detection Scheme}

In this framework, we consider a Li-ion battery pack with multiple modules where one or more than one modules can experience ISC. We assume that module voltages can be measured from the battery pack. Using the module voltage data we propose a KM-based algorithm to detect the shorted battery modules.



\begin{figure}[h!]
    \centering
    \includegraphics[width=0.8\linewidth]{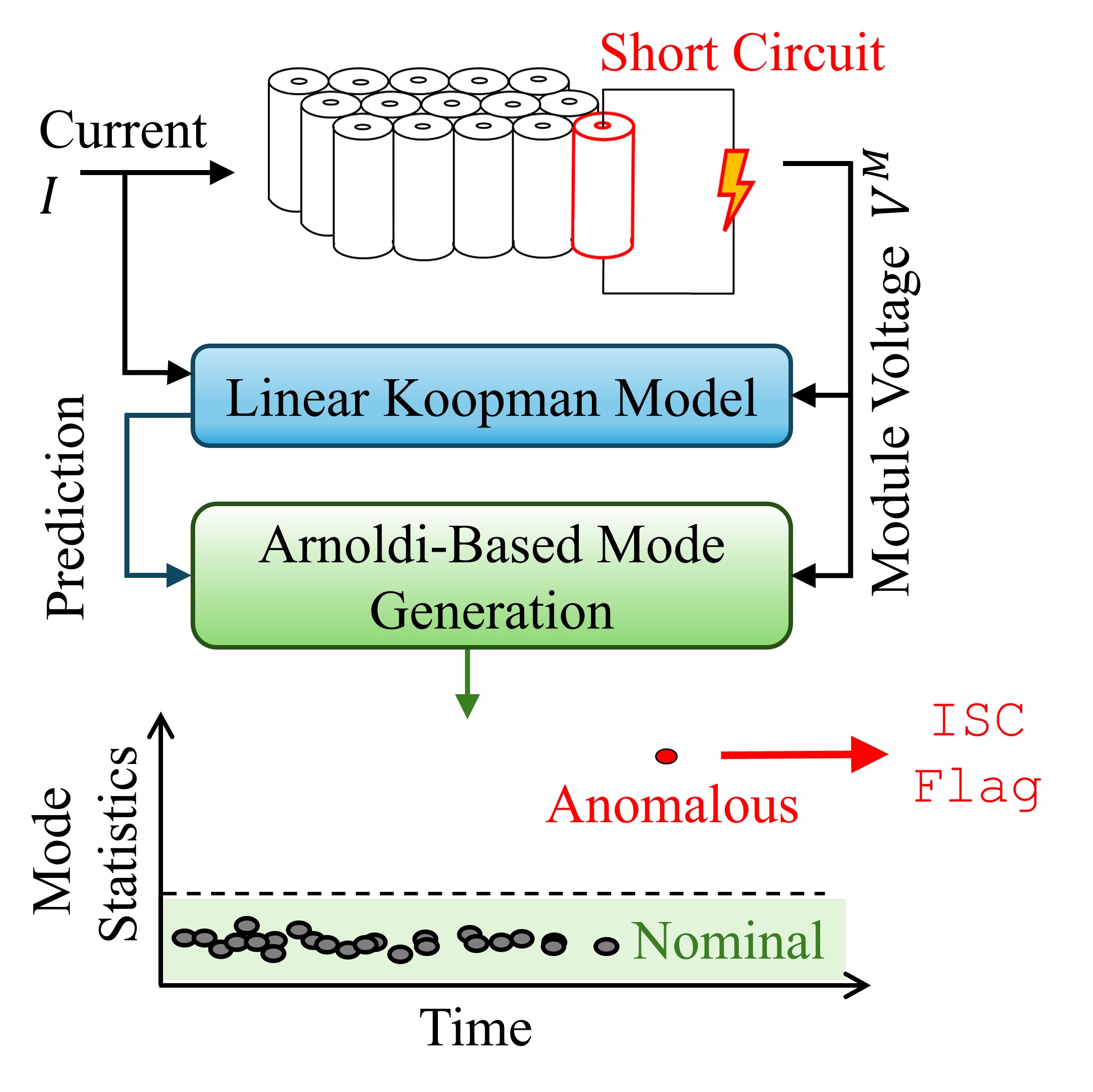}
    \caption{Block diagram showing the overview of the proposed ISC detection scheme for Lithium-ion battery packs.}
    \label{fig:detection}
\end{figure}

The KM-based ISC detection algorithm consists of three steps: (1) Obtaining the linear approximate Koopman model of each battery module from their module terminal voltage measurements using KMD, (2) Generation of KMs for each module using a second parallel KMD and utilizing the Arnoldi algorithm, and (3) Detection of shorted battery modules. Fig.~\ref{fig:detection} shows the block diagram of our proposed scheme by identifying statistical outliers among the modes of modules.


\textit{Linear approximate koopman model:} In this step, we utilize the module voltage time-series data over a receding horizon. First, we learn the approximate   Koopman linear model from the data over the learning window $\mathcal{L}$. Subsequently, we utilize the model to generate prediction $\hat{Y}_\mathcal{P}$ over the prediction window $\mathcal{P}$. Then, the sliding window is moved ahead with the length of the prediction window $\mathcal{P}$.

 \textit{Koopman mode generation:} KMs can capture the 
 changes in the behavior of the system \cite{ghosh2024koopman}. In particular, KMs can capture the inconsistencies both across the modules (spatial) and over time (temporal). We utilize these spatio-temporal discrepancies among various modules embedded in the KMs over time to detect the presence of ISC. Thus, we utilize the module voltage measurements $Y_\mathcal{P} $ over the prediction window $\mathcal{P}$ to  obtain the error sequence data $\mathcal{E}_s^i$ for the $i^{th}$ module among the $m$ number of modules as
 \begin{align}
     \mathcal{E}_s^i =  Y_\mathcal{P}^i - \hat{Y}_\mathcal{P}^i, \quad \forall i. \label{es}
 \end{align}
 Then, we deploy a second KMD scheme based on the Arnoldi algorithm to generate the sequence of KMs $\mathcal{E}_{M}^i$ for the error sequence data $\mathcal{E}_s^i$ for each module. 
 
 \textit{Detection of ISC:} The spatio-temporal discrepancy embedded in the KMs leads to the differences between the distribution of the KMs of a nominal module vs the KMs of a shorted module \cite{ghosh2024isolating}.
 We utilize kernel density estimation (KDE) \cite{botev2010kernel} based on a normal kernel function to estimate the distributions $\mathbf{P}^i$ of the KMs  for each module.
 Next, to measure and compare the differences among the KM distributions, we generate sample points  $z \in \mathcal{Z} = \begin{bmatrix}
     \underline{\mathcal{E}_{M}} & \overline{\mathcal{E}_{M}}
 \end{bmatrix}$ uniformly between $\underline{\mathcal{E}_{M}}$ and $\overline{\mathcal{E}_{M}}$ which are respectively the minimum and the maximum of KMs over all the modules. Then, we compute the Kullback-Leibler divergence (KLD) between two KM distributions   at these sample points \cite{kullback1951information}. The statistical distance KLD measures the difference  between two KM distributions and thus, captures the spatial or module-level discrepancies. The KLD is given by:
 \begin{align}
     { KLD(\mathbf{P}^i || \mathbf{P}^j) = \sum\limits_{z \in \mathcal{Z}} \mathbf{P}^i(z) log\left(\frac{\mathbf{P}^i(z)}{\mathbf{P}^j(z)}\right)}. \label{kld}
 \end{align}
 We note here that, $KLD(\mathbf{P}^i || \mathbf{P}^j) \neq KLD(\mathbf{P}^j || \mathbf{P}^i)$. Then, we compute the the average distance $\xi^i$ for each module as  
 \begin{align}
     \xi^i = {\sum\limits_j KLD(\mathbf{P}^i || \mathbf{P}^j)}\big/m, \quad \forall i, j . \label{kldAvg}
 \end{align}
To capture the temporal discrepancies embedded in the KMs, we evaluate the cumulative sum of the $\xi^i$ with time  and compare with minimum cumulatively summed $\xi^i$ for each module. Using this, we define our residual $r_k = \begin{bmatrix}
    r^1_k & \cdots & r^m_k
\end{bmatrix}$  at $k^{th}$ instant as:
\begin{align}
    r^i_k =  \Xi^i - \min\limits_i \left(\Xi^i\right),\,\,  \,\,  \Xi^i = \sum\limits_{k} \xi_k^i, \quad \forall i. \label{res}
  \end{align}

\begin{figure}[h!]
    \centering
    \includegraphics[width=0.9\linewidth]{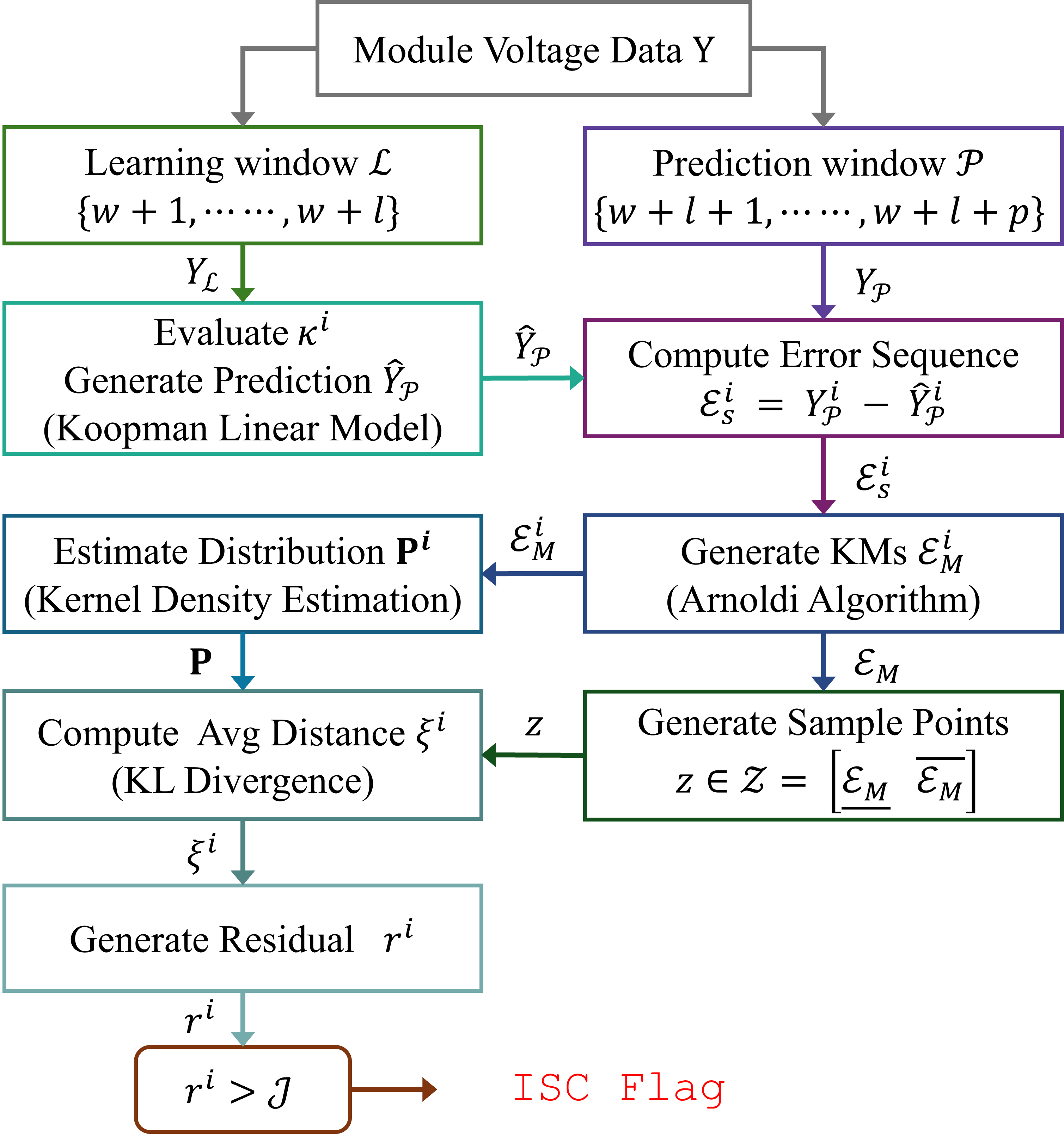}
    \caption{Flow chart showing the detailed steps for ISC flag generation using two parallel KMD schemes, Arnoldi algorithm, and KL divergence based statistical outlier detection, in case of ISC in battery modules. }
    \label{fig:algo}
\end{figure}

\textit{Choice of threshold:} The nominal average distance $\xi^i$ among the distributions of the KMs for the healthy modules may vary due to battery pack specifications, operating conditions, measurement noise, and sampling frequency. This leads to nominal fluctuations in the generated residual $r^i$, even in the absence of ISC. Hence, under nominal battery conditions, we use the fluctuations in the generated residuals initially to set the ISC detection threshold $\mathcal{J}$, representing the bound on these nominal fluctuations. Consequently, an ISC flag is generated for the $i^{th}$ module, if and when residual $r^i$ crosses this predefined threshold $\mathcal{J}$.  
%
Fig~\ref{fig:algo} shows the detailed steps of the proposed ISC detection scheme and Algorithm~\ref{alg:desicion} presents the implementation.

\begin{algorithm2e} \label{algo}
\caption{Generate \texttt{ISC Detection Flag},  \&    \texttt{Shorted Battery Module Index}}\label{alg:desicion}
\KwIn{Time instant $k$, module voltage measurements $y_k$,  learning window $\mathcal{L}$, prediction window $\mathcal{P}$, \& thresholds $\mathcal{J}$.}
\KwOut{ISC Flag with shorted  module Index.}
\SetKwFunction{FD}{Koopman  Model}
\SetKwFunction{FI}{KM Generator}
\SetKwFunction{FF}{ISC Detector}

    \For{$k\geqslant0$}{
    \For{i = 1:m}{
    $\hat{Y}_\mathcal{P}^i \quad \leftarrow$  \texttt{Koopman  Model($Y_\mathcal{L}^i,\mathcal{L}, \mathcal{P}$)}\\
    $\mathcal{E}_{M}^i \, \leftarrow$ \texttt{KM Generator(${Y}_\mathcal{P}^i , \hat{Y}_\mathcal{P}^i $)}\\
    $r^i \quad \,\,\, \leftarrow$ \texttt{ISC Detector($\mathcal{E}_{M}$)}\\
    \If{$r^i  \geqslant \mathcal{J}$}{
    \texttt{Shorted  Module Index} = i.\\
    \KwRet \texttt{ ISC  Flag} \;}}
    }

    \SetKwProg{Fn}{function}{:}{\KwRet}
    \Fn{\FD{$Y_\mathcal{L}^i,\mathcal{L}, \mathcal{P}$}}{
        Evaluate $\kappa^i$ using \eqref{win1}-\eqref{KOapp} and then find $\hat{Y}_\mathcal{P}^i$.\\
        \KwRet $\hat{Y}_\mathcal{P}^i$ \;
        }

    \SetKwProg{Fn}{function}{:}{\KwRet}
    \Fn{\FI{$\mathcal{E}_s^i$}}{
        Evaluate $\mathcal{E}_s^i$  using  \eqref{es} and then compute $\mathcal{E}_{M}^i$.\\
        \KwRet $\mathcal{E}_{M}^i$ \;
        }

    \SetKwProg{Fn}{function}{:}{\KwRet}
    \Fn{\FF{$\mathcal{E}_{M}$}}{
        Estimate the distributions $\mathbf{P}$  using KDE.\\
        Generate sample points  $z \in \mathcal{Z} = \begin{bmatrix} 
            \underline{\mathcal{E}_{M}} & \overline{\mathcal{E}_{M}}
        \end{bmatrix}$.\\
        Evaluate $\xi^i$ using \eqref{kld} -\eqref{kldAvg}.\\
        Compute residual $r^i$ using \eqref{res}.\\
        \KwRet $r_k$ \;
        }
\end{algorithm2e}

 
\textit{Generalizability of the proposed algorithm:} The proposed algorithm can be readily deployed to Li-ion battery packs of different chemistries without any pre-training with historical data since it does not require any specific knowledge such as pack configuration, battery cell chemistry, and battery aging characteristics. Furthermore, the algorithm does not rely on the   input current data and hence, remain effective for both resting and charging/discharging operating conditions of the battery packs. Moreover, the algorithm can learn the system dynamics in presence of measurement noise and thus, reliably detects any inconsistency in the measurement data due to the presence of ISC regardless of the uncertainties. 
\section{Simulation Results}
In this section, we presented the simulation case studies on shorted battery module detection for a Li-ion battery pack under both resting and charging conditions. Here, we considered a battery pack with 15 cells in a 3S5P configuration i.\,e., 5 modules connected in parallel and 3 cells connected in series in each module. The capacity of the pack was 25 $Ah$ and the nominal voltage was 10 $V$.  To generate the data, we adopted an Equivalent Circuit Model (ECM) of the battery pack using a set of coupled cell-level ECM models \cite{cheng2020optimal}. 
The dynamics of each battery cell in this pack is defined as:
\begin{align}
&\dot {SOC}_{ij} = \dfrac{-I_{ij}}{Q_{ij}}, \quad
{OCV}_{ij} = f(SOC_{ij}), \label{ocv} \\
&\dot V_{c,ij} = \dfrac{I_{ij}}{C_{ij}} - \dfrac{V_{c,ij}}{R_{c,ij}C_{ij}}, \label{vij}\\
&V_{ij} = OCV_{ij} - V_{c,ij} - I_{ij}R_{ij}. \label{Vij}
\end{align}
Here, $i \in \{1,2,\dots,5\}$ represents the module number and $j \in \{1,2,3\}$ represents the cell location within the module. $V_{ij}$ indicates the cell terminal voltage, $R_{ij}$ is the ohmic resistance, $R_{c,ij}$ is the  polarization resistance for the $(i,j)$ cell. Similarly, $I_{ij}$ is the current flowing through the cell, $C_{ij}$ is the ECM capacitance, $Q_{ij}$ is the cell capacity,  $SOC_{ij}$ is state-of-charge of the cell, and the cell open-circuit-voltage ${OCV}_{ij}$ is a function of ${SOC}_{ij}$. 
Then, using Kirchoff's current law,  the battery  pack current $I$ and the terminal voltage $V_t$ are given in terms of the module voltage $V^{i}$ and current $I^{i}$ as
$
   I = \sum_{i= 1}^5 I^{i}, \quad 
   V_t = V^{i} = \sum_{j=1}^{3} V_{ij}.
$
To generate the Koopman observables \eqref{win1}-\eqref{win2} for the HDMD method, we have considered the available module voltage measurements at $k^{th}$ time instant i.e. $y_k =  ~\begin{bmatrix}
    V^1_k & V^2_k & \cdots & V^5_k
\end{bmatrix}$. Additionally, we adopted the parameters of a $LiFePO_4$ cell from \cite{zhang2021multi} and are provided in Table \ref{table:Parameters}. Furthermore, we considered parameter uncertainties and measurement noise to demonstrate the efficacy of the proposed algorithm. We selected a learning window of 1500 data points and a prediction window of 700 data points for the algorithm. Additionally, we defined the threshold for detection at $\mathcal{J} = 1.2e9$. Under these experimental conditions, we conducted two case studies, and their results are presented below. The five modules are designated as M1, M2, M3, M4, and M5.
\begin{table}
\caption{Details On Battery Cell \& Experiment Parameters}
\centering
\begin{tabular}{|c| c |c| c |}
\hline
Cell Parameter  & Values & Experiment Parameter  & Values\\
\hline

Q & 5 $Ah$ & Parameter Uncertainty & 5\%  \\
C & 4.3 $kF$ & Sampling Rate & 100 Hz \\
R & 3.8 $m \Omega$ & Measurement Noise & $\pm$ 2 $m\,V$ \\
$R_c$ & 4 $m \Omega$ & SC Resistance & 15 $\Omega$ \\ 
\hline
\end{tabular}
\label{table:Parameters}
\end{table}

\subsection{ISC detection under resting condition} 
\begin{figure} [h!]
    \centering
    \includegraphics[width=1\linewidth]{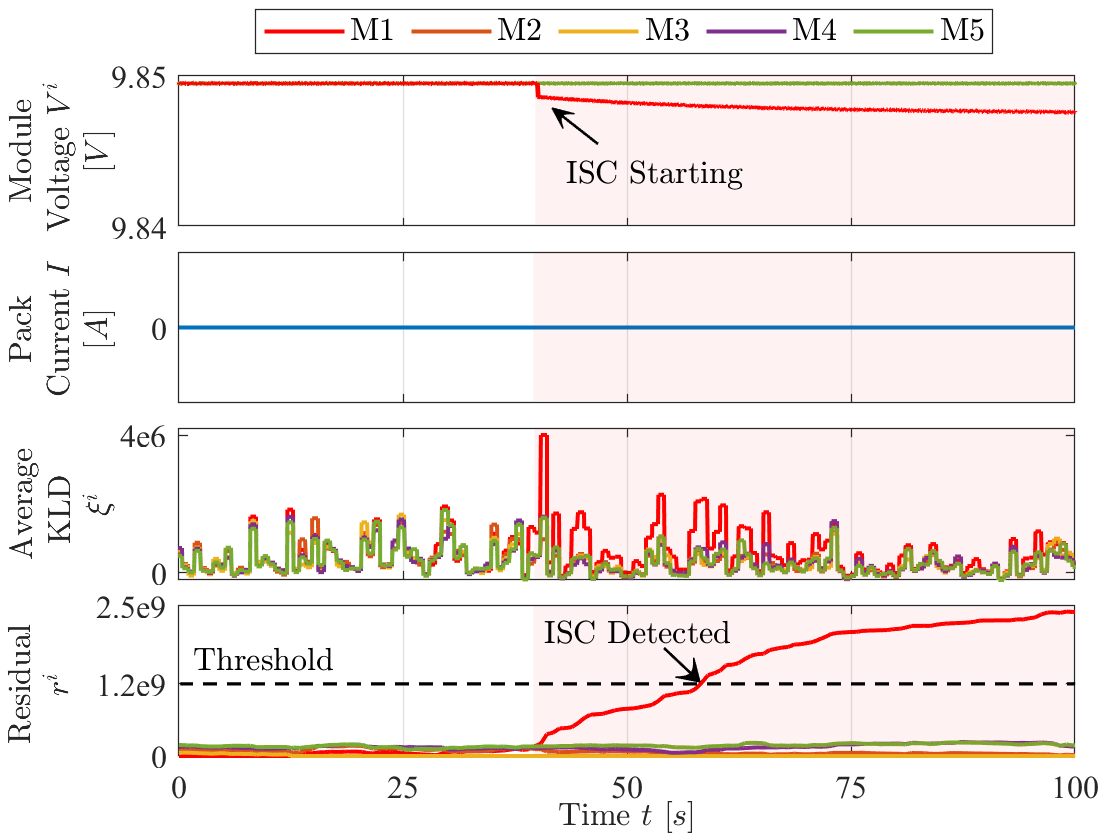}
    \caption{Plot shows (from top to bottom) the $i$-th module voltage, the pack current, the average distance among the distribution of KMs, and  the generated residual under resting condition.}
    \label{fig:enter-resting}
\end{figure}
For our first case study, we considered the battery pack at resting condition and the module-level voltages as measurements. Then, we introduced a soft ISC corresponding to a short circuit resistance of 15$\Omega$ such that the ISC affects one of the cells in the $1^{st}$ module from $30^{th}$ second. Such injection of ISC results in a drop in the module voltage measurement for the $1^{st}$ module. This scenario has shown on the top plot of Fig.~\ref{fig:enter-resting}. In the case of a resting battery, the pack current $I$ stays zero as shown in the $2^{nd}$ plot of Fig.~\ref{fig:enter-resting}. The $3^{rd}$ plot shows the computed average distances $\xi^i$ among the KM distributions for each module, and from the plot it is evident that the KM distribution of the shorted battery module diverged from the KM distribution of the healthy modules after ISC injection. Consequently, the generated residual $r^1$ crosses the threshold and accurately detects the presence of ISC at the $1^{st}$ module within 30s of ISC initiation. This has been shown in the last plot of Fig.~\ref{fig:enter-resting}.


\subsection{ISC detection under charging condition}
\begin{figure} [h!]
    \centering
    \includegraphics[width=1\linewidth]{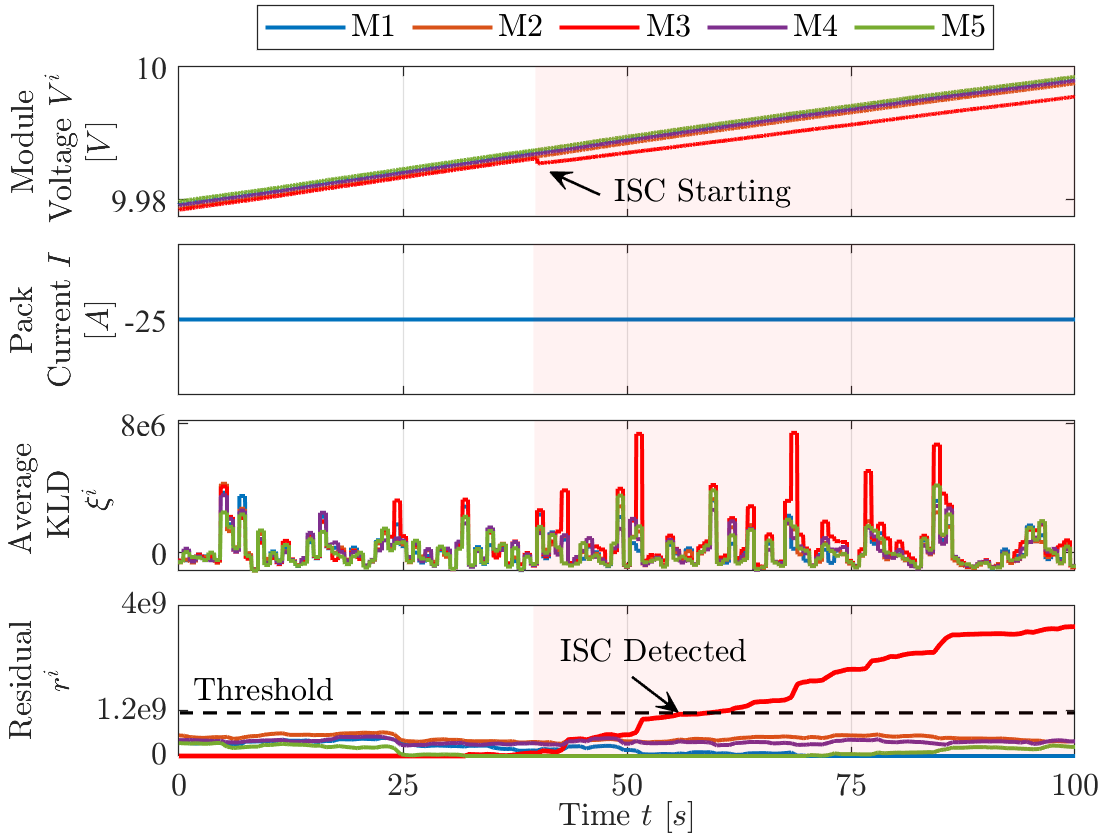}
    \caption{Plot shows (from top to bottom) the $i$-th module voltage, the pack current, the average distance among the distribution of KMs, and  the generated residual under constant current charging.}
    \label{fig:enter-charging}
\end{figure}

    For the second case study, we considered that the battery pack is under a constant charging condition since ISCs are often triggered during charging cycles wherein the charging current often masks the presence of ISC, especially for temperature-based detection strategies. We considered a $1C$ charging rate or a $25A$ charging current for our simulation. We injected ISC to one of the cells in $3^{rd}$ module at the $30^{th}$ second, using a short circuit resistance of 15$\Omega$. Consequently, a drop in voltage measurement occurred in the $3^{rd}$ module as shown in the top plot of Fig.~\ref{fig:enter-charging}. The $2^{nd}$ plot shows the pack current of $25A$ for this charging scenario. Here, the $3^{rd}$ plot captures the larger average distance for shorted module compared to the average distances of the healthy modules. The residual $r^3$ crossed the threshold to reliably generate the ISC flag within 30s of ISC injection as shown in the last plot of the Fig.~\ref{fig:enter-charging}.

    From our simulation results, we observe that in both the resting and charging conditions, the proposed algorithm accurately detected the presence of the ISC within 30s for the short circuit resistances of 15$\Omega$. In \cite{jia2024temperature}, authors utilized both voltage and temperature measurements along with the system model to detect ISC and for the short circuit resistances of 10$\Omega$ and 20$\Omega$, obtained a detection time of 73 min. In comparison, our proposed method yielded much faster detection while utilizing only the module voltage data. On the other hand, in \cite{cui2023internal}, authors adopted data-driven approaches to reduce detection time. However, their proposed algorithm required more than 75mins of ISC data for pre-training to successfully detect ISC with 10$\Omega$. Moreover, to reliably detect 15$\Omega$ ISC, they trained over even more historical data. Comparatively, our proposed algorithm is trained online with limited data over a 15s learning window. Therefore, our proposed algorithm exhibits the potential to significantly improve ISC detection benchmarks without the constraints of accurate battery pack models and reliance on extensive historical data. 
\section{Conclusions}\label{conclu}
In this paper, we have adopted a KM-based model-free approach to achieve faster detection of the ISC for a Li-ion battery pack. We have deployed the KO twice: first to learn the Koopman linear model for the battery modules and again to register the changes in the system through KMs generated based on the Arnoldi algorithm. Specifically, we have exploited the spatio-temporal discrepancies embedded in the KMs of various modules over time to capture the distinct dynamical behavior of the shorted modules. Thus, we estimate the distributions of the KMs for each module and monitor the differences among these KM distributions over time to locate the statistical outliers among the various KMs. Furthermore, the algorithm utilized only the module voltage data to identify the shorted battery modules. The simulation results illustrate the reliable and effectual detection performance of the algorithm under both resting and charging conditions. While accurate detection in presence of measurement noise and parameter uncertainties exhibits the robustness,  rapid detection under comparatively low severity ISC demonstrates the sensitivity of the algorithm. The performance of our algorithm motivates further performance evaluation with real-life experimental data and this will be covered in our future work. 





\bibstyle{arxiv}
\bibliography{ref.bib,Ref_Troy,Ref_Mallick}

\end{document}